\documentclass[aps,prl,twocolumn,superscriptaddress]{revtex4-2}
\usepackage{graphicx}
\usepackage{dcolumn}
\usepackage{bm}
\usepackage{color}
\usepackage{siunitx}
\usepackage{mathtools}
\usepackage{hyperref}
\usepackage{ulem}
\allowdisplaybreaks

\newcommand{\mrad}[1]{\SI{#1}{\milli\radian}}

\begin{document}

\preprint{}

\title{Coherent attosecond pulses generated by a relativistic electron beam interacting with an intense laser at a grazing angle}

\author{H. Peng}
\affiliation{Shenzhen Key Laboratory of Ultraintense Laser and Advanced Material Technology, Center for Intense Laser Application Technology, and College of Engineering Physics, Shenzhen Technology University, Shenzhen 518118, China}

\author{T.W. Huang}
\email{taiwu.huang@sztu.edu.cn}
\affiliation{Shenzhen Key Laboratory of Ultraintense Laser and Advanced Material Technology, Center for Intense Laser Application Technology, and College of Engineering Physics, Shenzhen Technology University, Shenzhen 518118, China}

\author{C.N. Wu}, 
\affiliation{Shenzhen Key Laboratory of Ultraintense Laser and Advanced Material Technology, Center for Intense Laser Application Technology, and College of Engineering Physics, Shenzhen Technology University, Shenzhen 518118, China}

\author{K. Jiang}
\affiliation{Shenzhen Key Laboratory of Ultraintense Laser and Advanced Material Technology, Center for Intense Laser Application Technology, and College of Engineering Physics, Shenzhen Technology University, Shenzhen 518118, China}

\author{R. Li}
\affiliation{Shenzhen Key Laboratory of Ultraintense Laser and Advanced Material Technology, Center for Intense Laser Application Technology, and College of Engineering Physics, Shenzhen Technology University, Shenzhen 518118, China}

\author{C. Riconda}
\affiliation{\mbox{LULI, Sorbonne Universit{\'e} CNRS {\'E}cole Polytechnique CEA, 75252 Paris, France}}

\author{S. Weber}
\affiliation{\mbox{Extreme Light Infrastructure ERIC, ELI-Beamlines Facility, 25241 Doln{\'\i} B{\u r}e{\u z}any,}\\ Czech Republic}

\author{C.T. Zhou}
\email{zcangtao@sztu.edu.cn}
\affiliation{Shenzhen Key Laboratory of Ultraintense Laser and Advanced Material Technology, Center for Intense Laser Application Technology, and College of Engineering Physics, Shenzhen Technology University, Shenzhen 518118, China}

\date{\today}

\begin{abstract}
  The interaction between relativistic electron beams and intense laser fields has been extensively studied for generating high-energy radiation. However, achieving coherent radiation from such interactions needs to precisely control the phase matching of the radiationg electrons, which has proven to be exceptionally challenging. In this study, we demonstrate that coherent attosecond radiation can be produced when a laser pulse interacts at grazing angle with a relativistic electron beam. The electrons oscillate in the laser field and are modulated with a superluminal phase, coherent ultrashort pulse trains are produced in the far field at the Cherenkov angle. This is verified by theoretical modeling and numerical simulations, including three-dimensional particle-in-cell (PIC) simulations and far-field time-domain radiation simulations. Based on our proposed scheme, high-repetition-rate, compact, and high-energy attosecond pulse sources are feasible.
  
  % The current laboratory conditions for lasers and relativistic electron beams meet the requirements of our proposed scheme, enabling its complete experimental realization.

\end{abstract}

\maketitle

Coherent attosecond pulses provide unprecedented temporal and spatial resolution, offering transformative potential for frontier research in atomic/molecular physics, nuclear physics, and ultrafast chemistry. These exceptional capabilities, recognized by the 2023 Nobel Prize in Physics\cite{krausz2009attosecond,krausz2014attosecond}, have opened new avenues for investigating electron dynamics and quantum processes at their natural timescales. Coherent attosecond pulses are routinely produced through gas-target high-harmonic generation (HHG)\cite{krausz2009attosecond,teubner2009high}, though limited by relatively low pulse energy and cutoff frequency. Relativistic laser-solid interactions \cite{lichters1996short,pukhov2010relativistic,dromey2012coherent} with ultrahigh temporal contrast offer a promising approach to enhance both attosecond pulse energy and cutoff frequency via plasma-surface HHG.

Coherent attosecond pulse generation via relativistic electron beam (REB)-laser interactions has recently attracted significant interest. Based on classical superradiance mechanism\cite{dicke1954coherence}, this usually requires the generation of nanoscale electron bunches to ensure phase matching on attosecond time-scale. Such REBs can be produced either by irradiating nanometer foils with high-contrast relativistic lasers\cite{wu2010uniform,wu2012giant} or through precisely controlled plasma wakefield acceleration using density-modulated downramps\cite{xu2022generation,xu2024attosecond} or abrupt upramp to plateau transition\cite{li2013dense,li2014radially}.  The second mechanism, generalized superradiance\cite{vieira2021generalized}, requires modulating a REB with superluminal phase velocity $v_m$ . This phase-locks radiating electrons at the Cherenkov angle $\theta_{ch}=\arccos(c/v_m)$ in the far field. A proposed proof-of-principle scheme employs a helical pre-modulated REB injected into a tightly-focused relativistic laser's focal plane. However, experimental realization remains challenging due to difficulties in preparing helical REBs and achieving precise spatiotemporal synchronization with the laser.

\begin{figure}[htbp]
  \centering
  \includegraphics[width=0.235\textwidth]{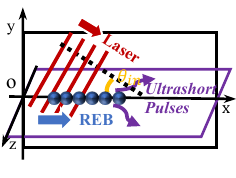}
  \includegraphics[width=0.235\textwidth]{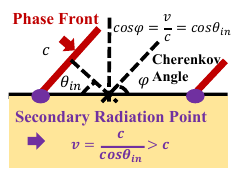}
  \caption{Left: scheme for coherent attosecond pulses generation from a REB interacting with an intense laser at a small angle $\theta_{in}$. Right: explaination of the light reflection from a surface with superluminal Cherenkov radiation mechanism. }
  \label{fig:scheme}
\end{figure}
 
In this letter, we propose a simple scheme to generate coherent attosecond pulses by exploiting the interaction between a REB and an intense laser at a grazing angle. As shown in the left of Fig.\ref{fig:scheme}, an s-polarized (z-direction) intense laser is injected obliquely towards a REB. The electrons propagate relativisticly along the x direction and undergo laser-driven oscillations in z. According to the classical radiation theory, each electron produces high-frequency radiation in the $xoz$ plane\cite{ride1995thomson,corde2013femtosecond}. Crucially, the laser phase velocity sweeping the REB is superluminal and the electrons slip back in the laser phase while performing phase-dependent oscillation. This impose a superluminal modulation on the collective REB motion, enabling coherent radiation at the Cherenkov angle with inherently short duration based on the generalized superradiance mechanism. (For clarity in explaining our scheme, here we model the laser as a plane wave. However, a Gaussian laser profile must be considered, including its phase evolution, as discussed later.)

This process is analagous to the light reflection at a flat surface. According to the Huygens-Fresnel principle, the reflected light is the constructive interference of the spherical waves emitted from the secondary radiation points, which are the crossing points of the light phase fronts and the surface. As shown in the right of Fig.\ref{fig:scheme}, as the light phase fronts propagate with the speed of light $c$ obliquely towards the surface with an injection angle of $\theta_{in}$, the secondary radiation point moves with a superluminal velocity $v=c/\cos\theta_{in}$ along the target surface. Then the secondary radiation point radiates coherently at the Cherenkov angle $\varphi$, forming the reflected light, which has an outgoing angle equal to the injection angle as $\cos\varphi=v/c=\cos\theta_{in}$.

Different from the light reflection, which happens at any incident angles, the interaction angle $\theta_{in}$ in our scheme needs to be small due to two facts: (1)the instaneous radiation from a relativistic electron is mainly contained within a small cone of opening angle of $\Delta\theta=1/\gamma_0$ centered on the propagation direction\cite{jackson1999classical,corde2013femtosecond}, where $\gamma_0$ is the mean relativistic factor of the REB. Thus $\theta_{ch}$ should be at least on the same order as $\Delta\theta$; (2) For a Gaussian laser, the laser waist $w_0$ is finite and the interaction length along the REB propagation direction can be estmimated as $L_{in}\approx 2w_0/\sin\theta_{in}$. Then with smaller interaction angle $\theta_{in}$, the interaction length is longer, which is benificial because the energy of the radiation pulses is proportional to the square of the interaction length for the superluminal coherent radiation mechanism\cite{vieira2021generalized,peng2023Coherent}. 

\begin{figure}[htbp]
  \centering
  \includegraphics[width=0.482\textwidth]{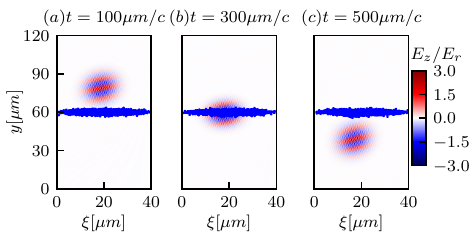}\\
  \includegraphics[width=0.482\textwidth]{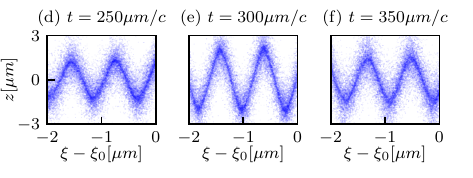}
  \caption{Schematic of simulations where the laser interacts with the REB at a grazing angle $\theta_{in}=\SI{100}{\milli\radian}$. Shown are the laser electric field and the electron bunch (scattered in blue). Note $E_r=m_ec\omega_0/e$, where $m_e$, $e$ are the electron charge and mass, $\omega_0$ is the laser frequency, respectively. $\xi=x-ct$ is the longitudinal coordinate in the comoving frame with speed $c$, while $\xi_0$ corresponding to the center of the moving window.}
  \label{fig:LaserAndEbunch}
\end{figure}

To validate the scheme mentioned above, the interaction between an intense laser and a REB at a small angle is simulated with the 3-dimensional particle-in-cell(PIC) code SMILEI\cite{DEROUILLAT2018351}, as shown in Fig.\ref{fig:LaserAndEbunch}. The cell numbers are $1600\times 1200 \times 600$($n_x\times n_y \times n_z$) at each direction, the spatial resolutions are $dx=\lambda_0/32=\SI{0.025}{\um}$, $dy=dz=\SI{0.1}{\um}$, where $\lambda_0=\SI{0.8}{\um}$ is the laser wavelength. The Gaussian laser is focused at $(x_{foc}=\SI{320}{\um}, L_y/2, L_z/2)$ with a laser waist $w_0=\SI{8}{\um}$, peak vector potential $a_0=3$ and temporal duration $\tau=\SI{20}{\fs}$, where $L_{x,y,z}$ is the simulation box size in each direction, respectively. The laser is s-polarized, i.e., linearly polarized in z direction. The REB is simulated with $\num{7d5}$ macroparticles with Gaussian random distribution in both longitudinal and transverse directions, the size along each direction are $\sigma_{x}=\SI{6}{\um}$ and $\sigma_{\perp}=\SI{1}{\um}$, respectively. The macroparticles have uniform weight and the total charge of the REB is about \SI{112}{\pico\coulomb}. For the low-current beam considered here, the self field of the REB and the space charge field are much smaller than the laser field and can be neglected\cite{ride1995thomson}, thus the electron macroparticles are set as test particles (without the laser the REB just transport ballistically towards the +x direction). The mean relativistic factor is $\gamma_0=100$ with a energy spread of \qty{1}{\percent} and the normalized emittance is \SI{100}{\nm~\radian}. Note that REBs with similar parameters and even better quality can be routinely produced with linear electron accelerators or plasma accelerators\cite{xiang2024ultrahigh,wang2021free,ke2021near}. The REB is also focused at $(x_{foc}=\SI{320}{\um}, L_y/2, L_z/2)$. The propagation direction of the laser has a small angle $\theta_l=\SI{-100}{\milli\radian}$ with the $+x$ axis. So the laser and REB interaction angle is $\theta_{in}=\SI{100}{\milli\radian}\approx\ang{5.73}$. The laser center and beam center overlap at the laser focus. A moving window with the speed $c$ is applied. 

As shown in Fig.\ref{fig:LaserAndEbunch}(a), at the beginning the REB is well separated from the laser to avoid unphysical sudden force on the electrons at $t=0$, the transverse distance between the laser center and the REB center is $D=x_{foc}\sin\theta_{in}\approx \SI{30}{\um}$ and is much larger than the laser beam size $w_{x0}=w_0\sqrt{1+(x_{foc}/x_R)^2}\approx \SI{12.5}{\um}$, where $x_R=\pi w_0^2/\lambda_0$ is the Rayleigh length of the laser. As they propagate towards each other as shown in Fig.\ref{fig:LaserAndEbunch}(b), the electron beam overlaps with the laser and the electrons are driven to oscillate along the z direction. They seperate from each other eventually. As shown in Fig.\ref{fig:LaserAndEbunch} and the supplemental movie, the REB envelope advances slowly in the comoving frame with the speed $c$, which means that the REB modulation phase is superluminal. The modulation phase velocity $v_{m}$ is estimated to be $v_{m}\approx 1.0025c$. The REB is supposed to produce attosecond pulses in the far field in the $xoz$ plane and the polarization is along $z$ direction according to the discussion above. 

\begin{figure}[htbp]
  \centering
  \includegraphics[width=0.482\textwidth]{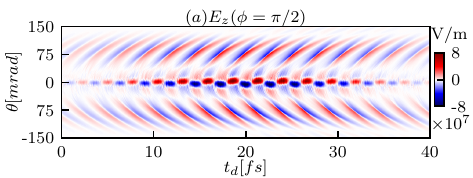}\\
  \includegraphics[width=0.482\textwidth]{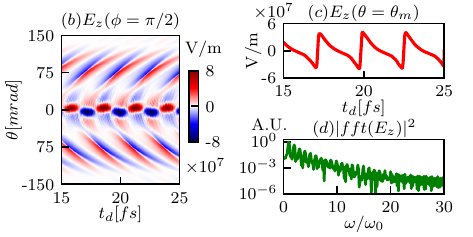}
  \caption{(a)The electrical field $E_z$ of the radiation on a far-field spherical detector plane with the azimuthal angle $\phi=\pi/2$. (b) The radiation emitted by the REB center. (c) The 1d temporal radiation field along the angle $\theta_m=\mrad{77.6}$. (d) The energy spectrum of the radiation field with arbitrary units (A.U.) shown in (c). }
  \label{fig:2dFarFieldRadiation}
\end{figure}

The far-field time-domain radiation is then computed by the post-processing code FaTiDo\cite{peng2023Coherent} using the trajectories of the macroparticles from SMILEI, as projected on a far-field detector, which is $r=\SI{0.1}{\m}$ away from the origin. The electric field of the radiation from every single macroparticle on the far-field detector is computed as 
\begin{gather}
  \mathbf{E}_{det}(\theta,\phi,t_d)=\frac{q}{4\pi\varepsilon_0}\frac{\mathbf{n}\times[(\mathbf{n}-\mathbf{\beta})\times\dot{\mathbf{\beta}}]}{cR(1-\mathbf{\beta}\cdot\mathbf{n})^3},
  \label{eq:farFieldRadiation}
\end{gather}
where $\theta$ is the polar angle with respects to the x axis, $\phi$ is the azimuthal angle between the observation direction with the $y$ axis in the $yz$ plane, $t_d=t_{ret}+\mathbf{R}/c$ is the detecting time, $t_{ret}$ is the retarded time and $\mathbf{R}$ is the vector from the macroparticle to the far-field detector at $t_{ret}$, respectively. Then the fields are interpolated on a evenly spaced detector time axis and summmed. The detecting time resolution is $dt_d=\SI{1e-18}{s}=\SI{1}{\as}$ and $N_{\theta}=800$ observers along the $\theta$ direction with $\theta$ ranging from $\SI{-150}{\milli\radian}$ to $\SI{150}{\milli\radian}$. 

As shown in Fig.\ref{fig:2dFarFieldRadiation}(a), off-axis intense coherent attosecond pulse trains can be found in the $xz$ plane with $\phi=\pi/2$. The radiation emitted by the REB center (around $t_d=\SI{20}{\fs}$) is much stronger and shorter, as shown in Fig.\ref{fig:2dFarFieldRadiation}(b),  exibiting a nonlinear dependence on the laser amplitude. The strongest and shortest off-axis radiation is found along the angle with $\theta_m\approx\SI{77.6}{\milli\radian}$, which is close to the Cherenkov angle $\theta_{ch}=\arccos(c/v_m)\approx \mrad{70.0}$ from theory. The one-dimensional temporal pulse profile is shown in Fig.\ref{fig:2dFarFieldRadiation}(c). The shortest duration(full width at half maximum, FWHM of the pulse envelope) of $E_z$ is about $\SI{391}{\as}$. The corresponding energy spectrum shown in Fig.\ref{fig:2dFarFieldRadiation}(d) is wide and extends to over $30\omega_0$. On the other hand, the REB envelope modulation induces microbunching at the laser period, producing periodic density variations along the propagation axis. Consequently, the REB emits attosecond trains of x-rays radiation primarily along the modulation direction $\theta\approx 0$\cite{hemsing2014beam,horny2020attosecond}, as shown in Fig.\ref{fig:2dFarFieldRadiation}(a, b). Another PIC simulation and the corresponding FaTiDo simulation with the laser copropagating with the REB, i.e. the interaction angle is $\theta_{in}=0$, is performed, no coherent attosecond pulses are found (not shown here), which proves well that the coherent attosecond pulse trains are indeed brought by the superluminal phase modulation.

The superluminal modulation of the REB by the Gaussian laser and the generation of coherent attosecond pulses can be explained as follows. In the coordinate system ($x_l, y_l, z_l$) where the laser propagates along the $+x_l$ axis and polarized along $z_l$ direction with the focal spot as the origin (note that $z_l$ direction is identical to the $z$ direction), the Gaussian laser phase is expressed as\cite{saleh2019fundamentals} 
\begin{gather}
  \varphi = kx_l-\omega t - \arctan{(x_l/x_R)} + \frac{kr_l^2}{2R}+ \varphi_0, \label{eq:GaussianLaserPhase}
\end{gather}
where $r_l^2 = y_l^2 + z_l^2$, $R=x_l\left[ 1+(x_R/x_l)^2\right]$ is the wavefront radius of curvature, respectively. The REB is injected at the angle $\theta_{in}$ with resepct to the laser central axis $+x_l$ in the $x_ly_l$ plane. Here, we focus on the regime of highly relativistic electrons ($\gamma_0\gg 1$) with weak laser perturbations ($a_0/\gamma_0\ll 1$), where the beam maintains its initial velocity $\mathbf{v}_0$ throughout the interaction, precluding laser-induced reflection or capture\cite{hartemann1995nonlinear,wang2001vacuum,salamin2002electronAcceleration,salamin2002electronscattering}. The phase velocity $\mathbf{v}_{\varphi}$ of a Gaussian laser along a particular trajectory $J$ is given by $\partial \varphi/\partial t+(\mathbf{v}_{\varphi})_J\cdot (\nabla \varphi)_J=0$\cite{pang2002subluminous}.
The phase velocity $\mathbf{v}_{\varphi}$ obtains the minimum value $v_{\varphi m}\approx c[1-1/(k_0w_0)^2]$ along the injection angle $\theta_m\approx 1/(k_0w_0)$, where $k_0=2\pi/\lambda_0$ is the laser wavevector. Along the injection direction $\theta_{in}$, the laser phase velocity can be estmimated as $v_{\varphi} \approx v_{\varphi m}/\cos(\theta_{in} - \theta_m)$, which can be superluminal with an injection angle $\theta_{in}\gtrsim 2.4/(k_0w_0)$, thus defining the minimum injection angle for superluminal modulation. During the interaction, the electrons are driven by the laser field and oscillate along the $z_l$ direction. The REB is modulated by the laser phase. The modulation velocity $v_m$ is equal to the laser phase velocity $v_{\varphi}$ along $\theta_{in}$, that is $v_m=v_{\varphi}(\theta_{in})$, thus it can be superluminal.

\begin{figure}[htbp]
  \centering
  \includegraphics[width=0.482\textwidth]{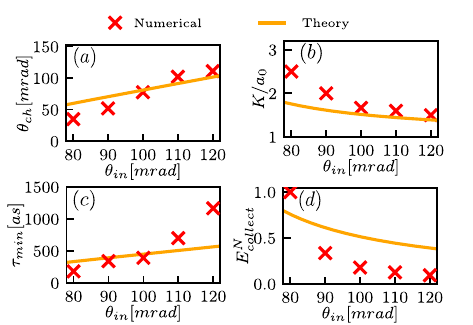}
  \caption{The dependence of (a) Cherenkov radiation angle $\theta_{ch}$, (b) the wiggler strength to laser amplitude ratio $K/a_0$, (c) minimum attosecond pulse duration $\tau_{min}$, (d) the normalized collective radiation field strength $E_{collect}^N$ (scaled to maximum) on the injection angle $\theta_{in}$, comparing theoretical predictions (orange curves) with simulation results (red crosses).} 
  \label{fig:scalingWithTheta_in}
\end{figure}

On the other hand, the single-electron oscillation in the laser field can be derived with the relativistic Newton-Lorentz equation of motion in the $z$ direction:
\begin{gather}
  \frac{dp_z}{dt} = -e(E_{zl}+v_{xl} B_{yl})\approx -e(1-\beta_0\cos\theta_{in})E_{zl},\label{eq:eqOfMotion}
\end{gather}
where $p_z=\gamma m_e v_z$ is the electron momentum in the $z$ direction, $v_{xl}$ is the electron velocity along $x_l$ direction, $E_{zl}$ and $B_{yl}\approx E_{zl}/c$ are the laser electric field and magnetic induction, respectively. Then the normalized electron momentum $u_z=p_z/(m_e c)=-\omega(1-\beta\cos\theta_{in}) \int a_z dt$, where $a_z=eE_z/(m_e\omega c)$ is the laser amplitude. For simplicity, we only consider the effect of the laser phase and neglect the transverse distribution of the laser amplitude, by setting $a_z=a_0 \cos\varphi$. As shown later, since all the electrons go through the laser transversely in the $y_l$ direction with a relatively small velocity $v_{yl}=\beta_0\sin\theta_{in}c$, this simplification works reasonablly well. We then have $u_z\approx -\omega_0(1-\beta_0\cos\theta_{in})a_0\sin\varphi/(d\varphi/dt)$, where $d\varphi/dt =\partial\varphi/\partial t+\mathbf{v}_0\cdot \nabla \varphi$ is the laser frequency experienced by the electron in the rest frame with velocity $\mathbf{v}_0$ and also the electron osicllation frequency in the lab frame $\omega_{\beta}$. Then we have $\omega_{\beta}=-d\varphi/dt=\omega_0(1-v_0/v_{\varphi})$ and $u_z\approx (1-\beta_0\cos\theta_{in})a_0\sin\varphi/(1-v_0/v_{\varphi})$. The wiggler strength of the electron oscillation $K=\gamma\psi=\gamma u_{zm}/u_{||}\approx u_{zm}$\cite{corde2013femtosecond}, where $u_{zm}$ is the maximum electron momentum in the $z$ direction, scales 
\begin{gather}
  K/a_0\approx u_{zm}/a_0\sim (1-\beta_0\cos\theta_{in})/(1-v_0/v_{\varphi}).
\end{gather}
Numerical evaluation of $K$ versus $\theta_{in}$ (parameters as in Fig.\ref{fig:LaserAndEbunch}) is shown in Fig.\ref{fig:scalingWithTheta_in}, which agrees well with PIC simulations. Notably, $K$ significantly exceeds $a_0$ for Gaussian beams and increases at smaller $\theta_{in}$, contrasting with the well-established plane-wave results where $K\sim a_0$ typically holds for electron-laser interaction. The discrepency between the theory and the PIC simulation at small angles can be explained by the fact that the electron oscillation amplitude is much larger and comparable to the laser waist, thus the transverse-uniformity approximation adopted-above is no longer valid. The analysis shows that the electrons oscillation is in the strong wiggler regime and the electrons emit off-axis light burtst centered on the $x$ axis within an opening angle $\Delta \theta\approx K/\gamma_0$. 

The superluminal modulation velocity $v_m$ combined with the wiggler-like off-axis emission from individual electrons, ensures coherent attosecond radiation at the Cherenkov angle. As theoretically demonstrated by J. Vieira et al.\cite{vieira2021generalized}, the collective radiation form factor $F(\omega, \theta)$ matches the single-electron spectrum at $\theta_{ch}=\cos^{-1}(c/v_m)$ resulting in an $N_e^2$ energy enhancement and ultrashort coherent radiation, where $N_e$ counts electrons swept by the laser phase. This form factor matching (occurring at the Cherenkov angle $\theta_{ch}$) directly manifests coherence and explains the observed attosecond pulses here. By relating the superluminal modulation velocity $v_m$ to the interaction angle $\theta_{in}$, a theoretical relationship between the Cherenkov angle $\theta_{ch}$ and injection angle $\theta_{in}$ is derived. As shown in Fig.\ref{fig:scalingWithTheta_in}, the simulated Cherenkov radiation angles agree well with theoretical predictions. The cutoff frequency and duration of the single-electron radiation can be estmimated as $\omega_c \approx 2n_c\gamma_0^2\omega_{\beta}/(1+K^2/2+\gamma_0^2\theta^2)$ and $\tau_c\approx 2\pi/\omega_c$, where $n_c\simeq 3K^3/4$ is the critical harmonic order and $\theta$ is the observation angle\cite{jackson1999classical}. At the Cherenkov angle, $\tau_c(\theta_{ch})$, is estmimated to be $\sim 10$s of attoseconds with minimal $\theta_{in}$ dependence. This results from competing effects: smaller $\theta_{in}$ enhances the wiggler strength $K$ while reducing $\theta_{ch}$, but simultaneously decreases the electron oscillation frequency $\omega_{\beta}$.

While analyzing REBs with realistic parameters, collective beam effects—such as finite beam waist, energy spread, and emittance—introduce additional phase shifts in single-electron radiation and broaden the collective radiation duration at the Cherenkov angle\cite{vieira2021generalized}. Specifically, the finite transverse beam size $\sigma_{\perp}$ results in a Gaussian distribution of arrival times for single-electron radiation bursts on the far-field detector, with a temporal spread of $\tau_b\sim \sigma_{\perp}\sin\theta_{ch}/c$. This spread, significantly longer than the single-electron burst duration, dictates the collective radiation duration. The theoretical estimate reproduces the simulation trend, as shown in Fig.\ref{fig:scalingWithTheta_in}(c).

The single-electron radiation power along the observation angle $\theta_{ch}$ is given by \cite{jackson1999classical} 
\begin{gather}
  \frac{dP(t_d)}{d\Omega}\approx \frac{2}{\pi}\frac{e^2}{m^2c^3}\gamma^4\left( \frac{dp_z}{dt}\right)^2\frac{(1-\gamma^2\theta_{ch}^2)^2}{(1+\gamma^2\theta_{ch}^2)^5},
\end{gather}
where the electric field scales as $E_{single}\propto \sqrt{\frac{dP}{d\Omega}}$. To estimate the collective radiation field strength $E_{collect}$, we must account for the coherent superposition of electrons swept by the superluminal laser phase and
the transverse distribution of the REB. During the interaction, electrons slip back in the laser phase by $l_{slip}=(v_{\varphi}-v_0)t_{in}$, where $t_{in}=2w_0/v_0\sin\theta_{in}$ is the transit time across the laser along  $y_l$. Incorporating the Gaussian overlap of single-electron radiation bursts, the collective field strength scales as $E_{collect}\propto (v_{\varphi}-v_0)E_{single}/(\sin\theta_{in}\sin\theta_{ch})$, which matches the simulation trend as shown in Fig.\ref{fig:scalingWithTheta_in}d. 

% The theoretical analysis reveals that reducing the angle is 

In summary, we have shown that an intense laser interacts with a relativistic beam at a small angle can introduce superluminal modulation on the beam, which then produces coherent attosecond pulses. A theoretical model is proposed to explain the superluminal modulation and the coherent radiation. This scheme is simple and highly robust, which may trigger further theoretical and experimental investigations. Especially, laser plasma wakefield accelerators are shown to be very efficient in generating high-quality ($\sim\SI{50}{\MeV}$) REBs\cite{wang2021free,ke2021near,xiang2024ultrahigh}. Combined with the laser plasma wakefield accelerators, our scheme can be made all-optical, high-repetition-rate and very compact. Moreover, our results can be generalized to the situation that both the driving laser and REB are much longer, e.g. for the REBs of picoseond-nanosecond durations from Linac, since the main physics are similar. Compared to the plasma-surface HHG, our scheme has no requirements on the laser contrast.

This work is supported by the National Key R\&D Program of China (Grants No. 2024YFA1613400 and No. 2022YFA1603300), the National Natural Science Foundation of China (Grants No. 12475248 and No. 12175154), the Shenzhen Science and Technology Program (Grant No. RCYX20221008092851073), the Natural Science Foundation of Guangdong (Grant No. 2025A1515012853), the Guangdong Province Key Construction Discipline Scientific Research Capacity Improvement Project (Grant No. 2021ZDJS107), and the Natural Science Foundation of Top Talent of SZTU (Grant No. GDRC202310).

\bibliography{references.bib}

\end{document}